\UseRawInputEncoding
\documentclass[aps,preprint]{revtex4}%
\usepackage{amsfonts}
\usepackage{amsmath}
\usepackage{amssymb}
\usepackage{multirow}
\usepackage{graphicx}
\usepackage{bm}%
\providecommand{\U}[1]{\protect\rule{.1in}{.1in}}

\begin{document}
\title{Mass spectra of doubly heavy tetraquarks in diquark-antidiquark picture$^{*}$ }
\author{YongXin Song$^{1}$}
\author{Duojie Jia$^{1,2}$\thanks{Corresponding author}}
\email{jiadj@nwnu.edu.cn}

\affiliation{$^{1}$Institute of Theoretical Physics, College of Physics and Electronic
Engineering, Northwest Normal University, Lanzhou 730070, China }
\affiliation{$^{2}$ Lanzhou Center for Theoretical Physics, Lanzhou University,
Lanzhou,730000,China }

\footnotetext{$^*$  Supported by the National Natural Science Foundation of China under the No. 12165017.}

\begin{abstract}
Inspired by recent observation of the first doubly charmed tetraquark $T_{cc}$, we apply linear Regge relation and mass scaling to study low-lying mass spectra of the doubly heavy tetraquark in heavy-diquark-light-antidiquark picture. The measured data and other compatible estimates of ground-state masses of doubly heavy baryons are employed to evaluate masses of heavy diquark $M_{QQ}$ ($Q=c,b$) and the $D/D_{s}$ meson masses are used in mass scaling to determine the hyperfine mass splitting. Our mass computation indicates that all low-lying states of doubly heavy tetraquarks are unstable against strong decays to two heavy-light mesons, except for the ground states of nonstrange $bb$ tetraquarks.
\end{abstract}

\maketitle

\section{Introduction}
Recently, the LHCb collaboration \cite{LHCb2021T} reported important observation of a doubly charmed tetraquark containing two charm quarks, an anti-u and an anti-d quark, using the LHCb-experiment data at CERN, which manifests itself as a narrow peak in the mass spectrum of $D^{0}D^{0}\pi^{+}$ mesons just below the $D^{*+}D^{0}$ mass threshold. This invite quantitative study of mass spectroscopy of the multiquark hadrons and rises issue as if there are more (strongly) stable doubly heavy tetraquark $T_{QQ}$ against the two-meson decay. Most of mass computations of the compact tetraquark $T_{cc}$ \cite{KR:pr2017,Eichten:pr2017,Luo:2017eub,Mehen:D17,M3H:D21,M3H:D22} predict masses around $3.9-4.1$ GeV, above the $D^{*+}D^{0}$ mass threshold ($3876$ MeV).

For doubly charmed hadrons, one crucial experimental input is the strength of the interaction between two charm quarks, which can be provided by the doubly charmed baryon $\Xi_{cc}^{++}=cuu$ discovered by the LHCb Collaboration at CERN. The updated mass of the doubly charmed baryon $\Xi_{cc}^{++}$ is $3621.55\pm 0.53$ MeV \cite{LHCb20J}. This value is consistent with several predictions, including the computed values of 3620 MeV \cite{EFG:D02} and $3627 \pm 12$ MeV \cite{KR:D2014}.

Interestingly, a narrow structure $X(6900)$ around $6.9$ GeV, the candidate
for all-charm tetraquark, was observed in 2020 in the $J/\psi $-pair mass
spectrum (above five standard deviations) by LHCb collaboration at CERN\cite{LHCb2020X69}, with the Breit-Wigner mass $m[X(6900)]=6905\pm 11\pm 7$ MeV and natural width $\Gamma [X(6900)]=80\pm 19\pm 33$ MeV. This observation is confirmed
recently by CMS experiment\cite{cms22} in the di-$J/\psi $ mass data with the
resonance mass $6927\pm 9\pm 5$ MeV. The recent ATLAS experiment at CERN
also finds a resonance in the di-$J/\psi $ channel \cite{atlas22}, with the mass $6.87\pm 0.03_{-0.01}^{+0.06}$ GeV and the width $0.12\pm
0.04_{-0.01}^{+0.03} $ GeV, which is also consistent with $X(6900)$ reported
by the LHCb. Some other broad structures at lower masses and around $6.9$
GeV are also seen in these experiments, whose natures are under study.

$M(cc\bar{q}\bar{q})$=$2M(ccu)$-$M(cc\bar{c}\bar{c})/2$ predicts mass of the tetraquark $T_{cc\bar{q}\bar{q}}$ to be around $3790.6\pm 10.06$ MeV, which is slightly below the $D^{*+}D^{0}$ mass threshold. Here, we used the mass $M(cc\bar{c}\bar{c})$=$6905\pm 18$ MeV, taken from the LHCb measured data of the fully charmed tetraquark $X(6900)$, around 6.9 GeV just above twice the $J/\Psi$ mass \cite{LHCb2020X69}.

We use Regge relation methods in Refs. \cite{Jia:2019bkr,Karliner:2015ema} and earlier works \cite{Song:2022csw}, which is validated by many successful prediction of the excited heavy baryons\cite{Jia:2019bkr,Karliner:2015ema}, to analyze low-lying mass spectra of the DH tetraquark. Masses of the DH tetraquark in their ground state and 1P wave are computed. In the analysis, the measured data of the $\Xi_{cc}^{++}$ \cite{LHCb20J} and other mass estimates of ground-state of doubly heavy baryons \cite{EFG:D02} that is compatible with this data are used to extract masses of heavy diquark $M_{QQ}$ ($Q=c,b$) and other trajectory parameters \cite{Song:2022csw}.

In the past, the doubly heavy tetraquarks in their ground state have been studied extensively in Refs. \cite{Ader:D82,Zouzou:C86,Manohar:B93,Bicudo:D15,Bicudo:D17,Mathur:D19S,LuCD:D20}, among many others. Some comprehesive review are given in Ref. \cite{ExoRev}. For recent review, see Refs. \cite{AliBk,Upd2022}.

\section{Method for Excited Spectrum}
We write mass of the doubly heavy (DH) tetraquarks $T_{QQ}$ as sum of two parts: $M=\bar{M}+\Delta M$, where $\bar{M}$ is the spin-independent part and
$\Delta M$ is the spin-dependent mass. In the picture of heavy-diquark-light-antidiquark($\bar{q}\bar{q}$), the
diquark $QQ$ consisting of two heavy quarks is quite heavy compared to light antidiquark $\bar{q}\bar{q}$ so that the heavy-light limit applies to
the DH tetraquarks $QQ\bar{q}\bar{q}$. One can then derive, by analogy with Regge
relation in Ref. \cite{Jia:2019bkr}, a linear Regge relation from the QCD string model for the DH tetraquark $T_{QQ}$ via viewing it to be a system of a massive QCD string with diquark $QQ$(D) at one end and $\bar{q}\bar{q}$($\bar{D}$) at the other.

\subsection{Spin-Independent Mass}

We consider systems of DH tetraquark $T_{QQ}$(=$QQ\bar{q}\bar{q}$) consisting of a S-wave heavy diquark $QQ$ and a S-wave light antidiquark $\bar{q}\bar{q}$, which are in relative states of the orbital angular momentum $L=0$ or $L=1$ between two diquarks. We also assume the heavy diquark $QQ$ are in color antitriplet ($\bar{3}_{c}$) and $\bar{q}\bar{q}$ in color triplet ($3_{c}$). Thus, the DH tetraquark we consider is that with color structure $(\bar{3}_{c}\times 3_{c})$. To describe $T_{QQ}$, especially its excited states, we employ linear Regge relation for the tetraquark $T_{QQ}$, which is rooted in general scattering theory of strong interaction and based on observed spectrum of established hadrons\cite{ColPR71}. The relation takes form \cite{Jia:2019bkr}
\begin{equation}\label{Regge}
\bar{M}_{L}=M_{Q Q}+\sqrt{\pi a L+\left[m_{qq}+M_{Q Q}-m_{\text {bare }Q Q}^{2} / M_{Q Q}\right]^{2}},
\end{equation}
where $M_{QQ}$ and $m_{qq}$ are the effective masses of the heavy and light diquarks, respectively, $a$ stands for the tension of the
QCD string connecting the diquark $QQ$ at one end and antidiquark $\bar{q}\bar{q}$ at the other. Here, $m_{\text{bare}QQ}$ is the bare mass of
$QQ$, given approximately by sum of the bare masses of each quark $Q$:
$m_{\text{bare }QQ}$$=m_{\text{bare}Q}+m_{\text{bare}Q}$, where $m_{bare,Q}$ of the quarks $Q$($=c,b$) are $m_{bare,c}=1.275$ GeV and $m_{bare,b}=4.18$ GeV\cite{PDG22}. Numerically, one has
\begin{equation}
m_{\text{bare }cc}=2.55~\text{GeV},m_{\text{bare }bb}=8.36\text{GeV}%
,m_{\text{bare }bc}=5.455\text{GeV}.   \label{mbare}%
\end{equation}
Applying to the S wave, one sees the reduction of the Regge relation (\ref{Regge}), $\bar{M}= M_{QQ}+m_{qq}+(M_{QQ}v_{QQ})^{2}/M_{QQ}$ where $v_{QQ}^{2}=1-(m_{\text{bare}QQ}/M_{QQ})^{2}$. It agrees with mass expansion at the heavy quark limit of DH hadrons.

In Table I, we list the relevant parameters in Eq. (\ref{Regge}) and their values. These values was evaluated previously in Refs. \cite{Jia:2019bkr,Song:2022csw,Jia:2020vek} via matching of the observed spectra of the singly-heavy hadrons.

\renewcommand{\tabcolsep}{0.08cm}
\renewcommand{\arraystretch}{1.5}
\begin{table}[!htbp]
\begin{center}
    \caption{Related parameters $(\rm GeV)$ of the QCD string model. Taken from Refs. \cite{Jia:2019bkr,Song:2022csw,Jia:2020vek}  }
    \label{tab:the effective mass}
    \begin{tabular}{ccccccccccc}
        \hline\hline
        {\rm Parameter} &{\rm $M_{cc}$} &{\rm $M_{bb}$}&{\rm $M_{\{b c\}}$}&{\rm $M_{\text {[bc}]}$}&{\rm $m_{\{ud\}}$}&
        {\rm $m_{\text {[ud}]}$}&{\rm $m_{\{us\}}$}&{\rm $m_{\text {[us}]}$}&{\rm $m_{\{ss\}}$}   \\ \hline
         The value        &2.8655             &8.9166    &5.8923     &5.8918     &0.745          &0.535        &0.872                   &0.718                 &0.991                    \\
        \hline\hline
    \end{tabular}
    \end{center}
\end{table}

(1) 1S wave. The relative orbital angular momentum $L=0$ with respect to $QQ$. In this case, the Regge relation (\ref{Regge}) becomes
\begin{align}
\bar{M}(1S)  &  =M_{QQ}+m_{qq}+\frac{k_{QQ}^{2}%
}{M_{QQ}},\label{ReggS}\\\nonumber
k_{QQ}^{2}  &  \equiv M_{QQ}^{2}-m_{\text{bare}QQ}%
^{2}.
\end{align}
Putting masses in Table \ref{tab:the effective mass} and Eq. (\ref{mbare}) into Eq.(\ref{ReggS}), one can obtain spin-averaged masses of all DH tetraquark $T_{QQ}$, as listed in Table \ref{tab:spin average}.

\renewcommand{\tabcolsep}{0.5cm}
\renewcommand{\arraystretch}{1.2}
\begin{table}
\begin{center}
    \caption{\rm Mean (spin-averaged) mass(in MeV) of the doubly heavy tetraquarks $T_{QQ}$ in ground states, $[]$ and $\{\}$ stands for $S_{[]}=0$ and $S_{\{\}}=1$.}
    \label{tab:spin average}
    \begin{tabular}{ccccc}
        \hline\hline
        {\multirow{2}{*}{\rm System}}   &{\rm State} & $QQ=cc$ & $QQ=bb$ & $ QQ=bc$\\\cline{2-5}
        & $IJ^{P}$ & \multicolumn{3}{c}{\rm Mass} \\ \hline
       $\{QQ\}[\bar{u}\bar{d}]$         & $01^{+}$     & $3997$    &10530  &7270   \\
       $[QQ][\bar{u}\bar{d}]$       & $00^{+}$     & $-$     &$-$    &7268   \\
       $[QQ]\{\bar{u}\bar{d}\}$       & $11^{+}$     & $-$     &$-$    &7478   \\\hline
       $\{QQ\}\{\bar{u}\bar{d}\}$ &$10^{+}$,$11^{+}$,$12^{+}$  & $4207$    &$10740$  &$7480$   \\\hline
        $\{QQ\}[\bar{u}\bar{s}]$         &$\frac{1}{2}1^{+}$     &4180    &10713  &7453   \\
       $[QQ][\bar{u}\bar{s}]$        & $\frac{1}{2}0^{+}$        & $-$     &$-$  &7451   \\
       $[QQ]\{\bar{u}\bar{s}\}$        & $\frac{1}{2}1^{+}$      & $-$     &$-$  &7605 \\\hline
       $\{QQ\}\{\bar{u}\bar{s}\}$ & $\frac{1}{2}0^{+}$,$\frac{1}{2}1^{+}$,$\frac{1}{2}2^{+}$    &$4334$     &$10867$  & $7607$  \\\hline
       {$\{QQ\}\{\bar{s}\bar{s}\}$}&  $00^{+}$,$01^{+}$,$02^{+}$  &$4450$    &$10987$  &$7725$   \\\hline
       $[QQ]\{\bar{s}\bar{s}\}$        &$01^{+}$     &$-$     &$-$  &7724   \\ \hline\hline
    \end{tabular}
    \end{center}
\end{table}

(2) 1P wave. As examined in Refs. \cite{Jia:2019bkr}, Regge slope $\pi a$ of heavy baryons, in the sense of quantum averaging in hadrons, is nearly independent of spin of lightdiquarks ($=0$ or $1$), but relies on the mass $M_{QQ}$ of heavy quarks. One then expects for tetraquark $T_{QQ}$ that its effective value of the tension $a$ varies only with the heavy content of the flavor combinations $QQ$.

Let us consider first the tetraquark $T_{QQ}=(QQ)(\bar{u}\bar{d})$ and scale its tension to the $\Lambda_{c,b}(=Qud, Q=c, b)$ baryons which
share the heavy-light structure similar to singly-heavy mesons. We assume, for simplicity, a power-law of the mass scaling for two string tensions of the doubly heavy tetraquark $(QQ)(\bar{u}\bar{d})$ and $\Lambda_{c, b}$:
\begin{equation}
\frac{a_{\Lambda_{c}}}{a_{\Lambda_{b}}}=\left(  \frac{M_{c}}{M_{b}}\right)  ^{P_{1}},
\label{Rel slope1}%
\end{equation}%
\begin{equation}
\frac{a_{\Lambda_{c}}}{a_{(QQ)(\bar{u}\bar{d})}}=\left(  \frac{M_{c}}{M_{QQ}%
}\right)  ^{P_{1}}, \label{slope}%
\end{equation}
Here, the relevant string tensions are evaluated previously in Ref \cite{Jia:2019bkr}, as listed in Table \ref{tab:string tension}.
\renewcommand{\tabcolsep}{0.1cm}
\renewcommand{\arraystretch}{1.2}
\begin{table}[!htbp]
\begin{center}
    \caption{The string tension coefficient $a$ $(\rm GeV)^{2}$ of $\Lambda_{c}/\Lambda_{b}$,$\Xi_{c}/\Xi_{b}$,$\Omega_{c}/\Omega_{b}$ baryons and the effective mass of component quark$(\rm GeV)$ of $M_{c}$ and $M_{b}$.}
    \label{tab:string tension}
    \begin{tabular}{cccccccccc}
        \hline\hline
        {\rm Parameters} &{\rm $M_{c}$} &{\rm $M_{b}$}&{\rm $m_{u}$}& {\rm $a_{\Lambda_{c}}$}&{\rm $a_{\Lambda_{b}}$}&{\rm $a_{\Xi_{c}}$}&{\rm $a_{\Xi_{b}}$}&{\rm $a_{\Omega_{c}}$}&{\rm $a_{\Omega_{b}}$}  \\ \hline
         The value        &1.44             &4.48  &0.23   &0.212     &0.246     &0.255          &0.307        &0.316                   &0.318                                 \\
        \hline\hline
    \end{tabular}
    \end{center}
\end{table}

The same procedure applies to the strange $T_{QQ}=(QQ)(\bar{u}\bar{s})$ and the associated $\Xi _{c,b}$($=c(us), b(us)$) baryons, the $T_{QQ}=(QQ)(\bar{s}\bar{s})$ and the $\Omega_{c,b}($=c(ss), b(ss)$)$ baryons, for which the mass scaling, corresponding to Eqs. (\ref{Rel slope1})
and (\ref{slope}), has the same form
\begin{equation}
\frac{a_{\Xi_{c}}}{a_{\Xi_{b}}}=\left(  \frac{M_{c}}{M_{b}}\right)  ^{P_{2}}, \frac{a_{\Xi_{c}}}{a_{(QQ)(\bar{u}\bar{s})}}=\left(  \frac{M_{c}}{M_{QQ}%
}\right)  ^{P_{2}},
\label{slope2}%
\end{equation}%

\begin{equation}
\frac{a_{\Omega_{c}}}{a_{\Omega_{b}}}=\left(  \frac{M_{c}}{M_{b}}\right)  ^{P_{3}}, \frac{a_{\Omega_{c}}}{a_{(QQ)(\bar{s}\bar{s})}}=\left(  \frac{M_{c}}{M_{QQ} }\right)  ^{P_{3}}.
\label{slope3}%
\end{equation}%

Putting the parameters in Table \ref{tab:string tension} into Eqs. (\ref{Rel slope1})-(\ref{slope3}), one obtains all power parameter $P$ and ensuing effective values of the string tension
\begin{equation}
{
\begin{array}
[c]{l}%
P_{1}=0.1311 \mathrm{\ }, P_{2}=0.1635 \mathrm{\ },
P_{3}=0.0056 \mathrm{\ },
\end{array}
}  \label{P}%
\end{equation}

\begin{equation}
\left\{
\begin{array}
[c]{l}%
a_{(cc)(\bar{u}\bar{d})}=0.2320 \mathrm{\ }{\rm GeV^{2}}, a_{(bb)(\bar{u}\bar{d})}=0.2692 \mathrm{\ }{\rm GeV^{2}}, a_{(bc)(\bar{u}\bar{d})}=0.2320 \mathrm{\ }{\rm GeV^{2}},\\
a_{(cc)(\bar{u}\bar{s})}=0.2854 \mathrm{\ }{\rm GeV^{2}}, a_{(bb)(\bar{u}\bar{s})}=0.3436 \mathrm{\ }{\rm GeV^{2}}, a_{(bc)(\bar{u}\bar{s})}=0.3211 \mathrm{\ }{\rm GeV^{2}},\\
a_{(cc)(\bar{s}\bar{s})}=0.3172 \mathrm{\ }{\rm GeV^{2}}, a_{(bb)(\bar{s}\bar{s})}=0.3192 \mathrm{\ }{\rm GeV^{2}}, a_{(bc)(\bar{s}\bar{s})}=0.3185 \mathrm{\ }{\rm GeV^{2}},
\end{array}
\right\}  \label{slope4}%
\end{equation}
Applying the data in Eq. (\ref{slope4}) and Table \ref{tab:the effective mass} to Eq. (\ref{Regge}), one can obtain the mean (spin-averaged) masses of the $T_{QQ}$ system $(QQ)(\bar{q}\bar{q})$ in P-wave($L=1$). The results are
\begin{equation}
\left\{
\begin{array}
[c]{l}%
\bar{M}_{(cc[\bar{u}\bar{d}])}  =4.283\mathrm{\ }\text{GeV, }\bar{M}_{(bb[\bar{u}\bar{d}])}  =10.774\mathrm{\ }\text{GeV,}\bar{M}_{(bc[\bar{u}\bar{d}])}  =7.535\mathrm{\ }\text{GeV,}\\
\bar{M}_{(cc\{\bar{u}\bar{d}\})}  =4.455\mathrm{\ }\text{GeV, }\bar{M}_{(bb\{\bar{u}\bar{d}\})}  =10.959\mathrm{\ }\text{GeV,}\bar{M}_{(bc\{\bar{u}\bar{d}\})}  =7.714\mathrm{\ }\text{GeV,}\\
\bar{M}_{(cc[\bar{u}\bar{s}])}  =4.485\mathrm{\ }\text{GeV, }\bar{M}_{(bb[\bar{u}\bar{s}])}  =10.992\mathrm{\ }\text{GeV,}\bar{M}_{(bc[\bar{u}\bar{s}])}  =7.748\mathrm{\ }\text{GeV,}\\
\bar{M}_{(cc\{\bar{u}\bar{s}\})}  =4.613\mathrm{\ }\text{GeV, }\bar{M}_{(bb\{\bar{u}\bar{s}\})}  =11.127\mathrm{\ }\text{GeV,}\bar{M}_{(bc\{\bar{u}\bar{s}\})}  =7.879\mathrm{\ }\text{GeV,}\\
\bar{M}_{(cc\{\bar{s}\bar{s}\})}  =4.741\mathrm{\ }\text{GeV, }\bar{M}_{(bb\{\bar{s}\bar{s}\})}  =11.216\mathrm{\ }\text{GeV,}\bar{M}_{(bc\{\bar{s}\bar{s}\})}  =7.981\mathrm{\ }\text{GeV.}
\end{array}
\right\}  \label{hdiquark Xi}%
\end{equation}
where $QQ=cc, bb, bc$ stand for the axial diquark $\{QQ\}$ with spin $S_{\{QQ\}}=1$.

\subsection{Spin-Dependent Mass}

For mass splitting $\Delta M=\langle H^{SD}\rangle$ due to spin interaction between heavy and light diquarks with spins $\mathbf{S}_{QQ}$ and  $\mathbf{S}_{qq}$, we consider the spin-dependent Hamiltonian \cite{Karliner:2015ema,Ebert:2011jc}%
\begin{equation}%
\begin{array}
[c]{c}%
H^{SD}=a_{1}\mathbf{L}\cdot\mathbf{S}_{qq}+a_{2}\mathbf{L}\cdot\mathbf{S}%
_{QQ}+bS_{12}+c\mathbf{S}_{qq}\cdot\mathbf{S}_{QQ},\\
S_{12}=3\mathbf{S}_{qq}\cdot\hat{\mathbf{r}}\mathbf{S}_{QQ}\cdot
\hat{\mathbf{r}}-\mathbf{S}_{qq}\cdot\mathbf{S}_{QQ},
\end{array}
\label{spin-d}%
\end{equation}
where the first two terms are spin-orbit forces, the third is a tensor force,
and the last describes hyperfine splitting. We discuss the following states of the DH tetraquark:

1)1S wave. For the 1S wave, $L=0$, the spin-interaction Hamiltonian is simply
\begin{equation}
H^{SD}=c\mathbf{S}_{QQ}\cdot\mathbf{S}_{qq}\text{,} \label{SD1}%
\end{equation}
in which $\mathbf{S}_{QQ}\cdot\mathbf{S}_{qq}$ has the eigenvalues
${-2,-1,1,0}$ when $\mathbf{S}_{QQ}=1$. One has mass formula then,
\begin{equation}
M(QQ\bar{q}\bar{q},1S)=\bar{M}(QQ\bar{q}\bar{q})+c(QQ\bar{q}\bar{q})diag\{-2,-1,1,0\}, \label{mass1}%
\end{equation}
Based on the similarity between $T_{QQ}$ and heavy mesons $Q\bar{q}$(we choose $D$ meson typically), one has a relation of mass scaling for the spin coupling c,
\begin{equation}
c{(\{QQ\}(\bar{q}\bar{q}))}=\frac{M_{c}}{M_{QQ}}\cdot\frac{m_{q}}{m_{qq}}\cdot c(D)_{1S}  . \label{C1}%
\end{equation}%
in which $c(D)_{1S}=140.6$ MeV, $m_{q}=230$ MeV and others masses are given in Table \ref{tab:the effective mass} and Table \ref{tab:string tension}.

(2) 1P wave. In the heavy-diquark(D)-light-antidiquark($\rm \bar {D}$) picture, the total spin of the $T_{QQ}$ is denoted by
$S_{tot}=S_{D}+S_{\bar {D}}$, which takes value $S_{tot}={2,1,0}$ when $S_{D}=S_{\bar {D}}=1$ and $S_{tot}=1$ when $S_{D}=1$ and $S_{\bar {D}}=0$. In the scheme of $LS$ coupling, coupling $S_{tot}=2$ to $L=1$ gives the states with the total angular momentums $J=3,2,1$, while coupling
$S_{tot}=1$ to $L=1$ leads to the states with $J=2,1,0$; Coupling
$S_{tot}=0$ to $L=1$ leads to the states with $J=1$. Normally, one uses the $LS$ basis $^{2S_{tot}+1}P_{J}$ $=\{^{3}P_{0}%
,^{1}P_{1},^{3}P_{1},^{5}P_{1},^{3}P_{2},^{5}P_{2},^{5}P_{3}\}$ to label these multiplets in P-wave. We consider two cases for DH tetraquarks:

(a)$S_{D}=S_{\bar {D}}=1$. There are seven states, all of which are negative parity. In the $LS$ basis, the three $J=1$ states and two $J=2$ states are unmixed unless $a_{1}=a_{2}$. Otherwise
they are the respective eigenstates of a $3\times3$ and $2\times2$ matrices $\Delta {M}_{J}$ representing $H^{SD}$ with $J=1$ and $J=2$. The matrices of mass shifts are then(see Appendix A).
\begin{equation} \label{J=0}
\Delta {M}_{J=0}=-a_{1}-a_{2}-2b-c,
\end{equation}
\begin{equation}
\begin{aligned}   \label{J=1}
\Delta M_{J=1} &=\left[\begin{array}{ccc}
0 & \frac{2}{\sqrt{3}}\left(a_{1}-a_{2}\right) & 0 \\
\frac{2}{\sqrt{3}}\left(a_{1}-a_{2}\right) & \frac{1}{2}\left(a_{1}+a_{2}\right) & \frac{\sqrt{5}}{6}\left(a_{1}-a_{2}\right) \\
0 & \frac{\sqrt{5}}{6}\left(a_{1}-a_{2}\right) & -\frac{3}{2}\left(a_{1}+a_{2}\right)
\end{array}\right], \\
&+b\left[\begin{array}{ccc}
0 & 0 & \frac{32}{15 \sqrt{5}} \\
0 & 1 & 0 \\
\frac{32}{15 \sqrt{5}} & 0 & -\frac{7}{5}
\end{array}\right]+c\left[\begin{array}{ccc}
-2 & 0 & \frac{1}{6 \sqrt{5}} \\
0 & -1 & 0 \\
\frac{1}{6 \sqrt{5}} & 0 & 1
\end{array}\right],
\end{aligned}
\end{equation}
\begin{align}   \label{J=2}
\Delta M_{J=2}=\left[\begin{array}{ll}
\frac{1}{2}\left(a_{2}+a_{1}\right) & \frac{\sqrt{3}}{2}\left(a_{1}-a_{2}\right) \\
\frac{\sqrt{3}}{2}\left(a_{1}-a_{2}\right) & \frac{1}{2}\left(a_{2}+a_{1}\right)
\end{array}\right]+b\left[\begin{array}{cc}
-\frac{1}{5} & 0 \\
0 & \frac{7}{5} \\
\end{array}\right]+c\left[\begin{array}{cc}
-1 & 0 \\
0 & 1
\end{array}\right],
\end{align}
\begin{equation}  \label{J=3}
\Delta {M}_{J=3}=a_{1}+a_{2}-\frac{2}{5}b+c,
\end{equation}

In the heavy quark limit, the terms involving $\mathbf{S}_{QQ}$ in Eq. (13) behave as $1/M_{QQ}$ and are suppressed. Due to heavy quark spin symmetry($\mathbf{S}_{QQ}$ conserved), the total angular momentum of
the light quark $\mathbf{j}=\mathbf{L}+\mathbf{S}_{qq}\mathbf{=J-S}
_{QQ}$ is conserved and forms a set of the conserved operators
\{$\mathbf{J},\mathbf{j}$\}, together with the total angular momentum
$\mathbf{J}$ of the DH tetraquark. So instead of the $LS$ coupling, one can use the $jj$ coupling scheme in which the light-antidiquark spin
$\mathbf{S}_{qq}$ couples to $\mathbf{L}$ to form total angular momentum
$\mathbf{j}$ of the light quark and then to the conserved spin $\mathbf{S}_{QQ}$ of the heavy-diquark. So, for the tetraquarks $T_{QQ}$, we label the
P-wave multiplets in terms of the basis $|J, j\rangle$ of the $jj$ coupling (Appendix A), in which the diquark $QQ$ is infinitely heavy ($M_{QQ}\rightarrow\infty$) and $\mathbf{L}\cdot\mathbf{S}_{qq}$ becomes diagonal. As
such, one can employ the $jj$ coulpling to find the formula for mass spliting
$\Delta M$ by diagonalizing $\mathbf{L}\cdot\mathbf{S}_{qq}$ and treating other
interactions in Eq. (\ref{spin-d}) proportional to $a_{2}$, $b$ and $c$ perturbatively.
The results for $\Delta M(J,j)$ are listed in Table \ref{tab:matrix elements}.
\renewcommand{\tabcolsep}{0.45cm} \renewcommand{\arraystretch}{1.2}
\begin{table}[tbh]
\begin{center}
\caption{The matrix elements of the mass splitting operators in the P-wave $T_{QQ}$ states in the $jj$ coupling.}%
\label{tab:matrix elements}
\begin{tabular}
[c]{ccccc}\hline\hline
\textrm{(J, j)} & \textrm{$\left\langle \mathbf{L}\cdot\mathbf{S}_{qq}\right\rangle $}& \textrm{$\left\langle \mathbf{L}\cdot\mathbf{S}_{QQ}\right\rangle $} & \textrm{$\left\langle \mathbf{S}_{12}\right\rangle $} &
\textrm{$\left\langle \mathbf{S}_{qq}\cdot\mathbf{S}_{QQ}\right\rangle
$}\\\hline
$(0,0)$ & $-1$& $-1$   & $-2$      & $-1$\\ \hline
$(1,0)$ & $-2$& $0$    & $4/135$   & $1/27$\\
$(1,1)$ & $-1$& $-1/2$ & $-47/45$  & $-5/9$\\
$(1,2)$ & $1$ & $-2/3$ & $83/135$  & $-40/27$\\\hline
$(2,1)$ & $-1$& $1/2$  & $1$       & $1/2$\\
$(2,2)$ & $1$ & $-1/2$ & $1/5$     & $-1/2$\\\hline
$(3,2)$ & $1$ & $1$    & $-2/5$    & $1$\\\hline\hline
\end{tabular}
\end{center}
\end{table}
Given the matrices in Table 4, one can use the lowest perturbation theory to find the mass spliting $\Delta M(J,j)$ of the DH tetraquarks in P-wave. The result is
\begin{equation}
\Delta {M}(0,0)=-a_{1}-a_{2}-2b-c, \label{0.0}%
\end{equation}
\begin{equation}  \label{1.0}
\Delta {M}(1,0)=-2a_{2}+\frac{4}{135}b+\frac{1}{27}c,
\end{equation}
\begin{equation}  \label{1.1}
\Delta {M}(1,1)=-a_{1}-\frac{1}{2}a_{2}-\frac{47}{45}b-\frac{5}{9}c,
\end{equation}
\begin{equation}  \label{1.2}
\Delta {M}(1,2)=a_{1}-\frac{3}{2}a_{2}+\frac{83}{135}b-\frac{40}{27}c,
\end{equation}
\begin{equation}  \label{2.1}
\Delta {M}(2,1)=-a_{1}+\frac{1}{2}a_{2}+b+\frac{1}{2}c,
\end{equation}
\begin{equation}  \label{2.2}
\Delta {M}(2,2)=a_{1}-\frac{1}{2}a_{2}+\frac{1}{5}b-\frac{1}{2}c,
\end{equation}
\begin{equation}  \label{3.2}
\Delta {M}(3,2)=a_{1}+a_{2}-\frac{2}{5}b+c,
\end{equation}
which express the $T_{QQ}$ mass splitting in terms of four parameters($a_{1},a_{2},b,c$) of the spin couplings. Adding the spin-dependent
$\bar{M}=\bar{M}(1P)$, one obtains the respective mass formula $M(J,j)=\bar{M}(1P)+\Delta M(J,j)$ for the $T_{QQ}$ in P-wave. Here, one can check that the spin-weighted sum of these mass shifts is zero:
\begin{equation}
\sum_{J}(2J+1)\Delta M(J,j)=0. \label{sumJ}%
\end{equation}

To evaluate spin coupling parameters ($a_{1},a_{2},b,c$) in Eq.~(\ref{0.0})
through Eq.~(\ref{3.2}), we use the following mass scaling,
\begin{equation}  \label{aa1}
a_{1}[Q Q(\bar{q}\bar{q})]=\frac{M_{c}}{M_{Q Q}} \cdot \frac{m_{s}}{m_{(q q)}} \cdot a_{1}\left(D_{s}\right),
\end{equation}
\begin{equation}  \label{aa2}
a_{2}[Q Q(\bar{q}\bar{q})]=\frac{M_{c}}{M_{Q Q}} \cdot \frac{m_{s}}{1+m_{(q q)/M_{c}}} \cdot a_{2}\left(D_{s}\right),
\end{equation}
\begin{equation}  \label{b1}
b[Q Q(\bar{q}\bar{q})]=\frac{M_{c}}{M_{Q Q}} \cdot \frac{m_{s}}{1+m_{(q q)/m_{s}}} \cdot b\left(D_{s}\right),
\end{equation}
\begin{equation}  \label{c1}
c[Q Q(\bar{q}\bar{q})]=\frac{M_{c}}{M_{Q Q}} \cdot \frac{m_{ss}}{m_{(q q)}} \cdot c\left(\Omega_{c}\right),
\end{equation}
which apply  successfully to the singly-heavy baryons \cite{Jia:2019bkr,Karliner:2015ema}. For this, we list the parameters of spin couplings for the
$D_{s}$ and $\Omega_{c}$ in Refs. \cite{Jia:2019bkr,Karliner:2015ema,Jia:2020vek} and the effective masses of $m_{ss}$ and $m_{s}$ in Refs. \cite{Jia:2019bkr,Jia:2020vek} collectively in Eq. (\ref{spinEf}).
\begin{equation}
\left\{
\begin{array}
[c]{l}%
a_{1}(D_{s})=89.36  \mathrm{\ }{\rm MeV},
a_{2}(D_{s})=40.7 \mathrm{\ }{\rm MeV},
b(D_{s})=65.6 \mathrm{\ }{\rm MeV},\\
c(\Omega_{c})=4.04 \mathrm{\ }{\rm MeV},
m_{s}=328 \mathrm{\ }{\rm MeV},
m_{ss}=991 \mathrm{\ }{\rm MeV},
\end{array}
\right\}  \label{spinEf}%
\end{equation}
Putting the data in Eq. (\ref{spinEf}) and Tables \ref{tab:string tension} and \ref{tab:the effective mass} into Eqs.~(\ref{aa1})-(\ref{c1}), one obtains the spin couplings for the doubly heavy tetraquarks $T_{QQ}$, with the results listed collectively in Table \ref{tab:$T_{QQ}$}.

\renewcommand{\tabcolsep}{0.25cm}
\renewcommand{\arraystretch}{1.2}
\begin{table}[!htbp]
\begin{center}
\caption{Spin-coupling parameters (in MeV) in the spin interaction
(\ref{spin-d}) of the doubly heavy tetraquark $T_{QQ}$ with vector diquark $\{\bar{n}\bar{n}\}$.  }%
\label{tab:$T_{QQ}$}
\begin{tabular}
[c]{cccccccccc}\hline\hline
$T_{QQ}$ & $cc\{\bar {u}\bar {d}\}$ & $bb\{\bar {u}\bar {d}\}$ & $bc\{\bar {u}\bar {d}\}$
         & $cc\{\bar {u}\bar {s}\}$ & $bb\{\bar {u}\bar {s}\}$ & $bc\{\bar {u}\bar {s}\}$
         & $cc\{\bar {s}\bar {s}\}$ & $bb\{\bar {s}\bar {s}\}$ & $bc\{\bar {s}\bar {s}\}$  \\ \hline
\textrm{$a_{1}$} & $19.77$   & $6.35$   & $9.61$ & $16.89$  & $5.43$   & $8.21$   & $14.86$ & $4.78$  & $7.23$\\
\textrm{$a_{2}$} & $13.48$   & $4.33$   & $6.56$ & $12.74$  & $4.09$   & $6.20$   & $12.12$ & $3.89$  & $5.89$\\
\textrm{$b$}     & $10.08$   & $3.42$   & $4.90$ & $9.01$   & $2.90$   & $4.38$   & $8.20$  & $2.63$  & $3.99$\\
\textrm{$c$}     & $2.70$    & $0.87$   & $1.31$ & $2.31$   & $0.74$   & $1.12$   & $2.03$  & $0.65$  & $0.99$\\ \hline\hline
\end{tabular}
\end{center}
\end{table}

(b)$S_{D}=1$,$S_{\bar {D}}=0$. There are three states with $J=0,1,2$, corresponding to mass shifts of the spin interaction as follow
\begin{equation}  \label{a11}
\Delta M_{J=0}=-2a_{2},
\end{equation}
\begin{equation}  \label{a22}
\Delta M_{J=1}=-a_{2},
\end{equation}
\begin{equation}  \label{a33}
\Delta M_{J=2}=a_{2},
\end{equation}
for which $\sum_{J}(2J+1)\Delta M_{J}=0$ (Appendix A). Using the data in Eq. (\ref{spinEf}) and Tables \ref{tab:string tension} and \ref{tab:the effective mass},  Eq.~(\ref{aa2}) enables us to compute the spin coupling $a_{2}$, listed as below:
\begin{equation}
\left\{
\begin{array}
[c]{l}%
a_{2}(cc[\bar{u}\bar{d}])=14.91  \mathrm{\ }{\rm MeV},
a_{2}(bb[\bar{u}\bar{d}])=4.79 \mathrm{\ }{\rm MeV},
a_{2}(bc[\bar{u}\bar{d}])=7.25 \mathrm{\ }{\rm MeV},\\
a_{2}(cc[\bar{u}\bar{s}])=13.65 \mathrm{\ }{\rm MeV},
a_{2}(bb[\bar{u}\bar{s}])=4.39 \mathrm{\ }{\rm MeV},
a_{2}(bc[\bar{u}\bar{s}])=6.64 \mathrm{\ }{\rm MeV},
\end{array}
\right\}  \label{parameter}%
\end{equation}

\section{Numerical Results and Discussions}

Based on Section 2.2, one can calculate spin-dependent mass $\Delta M$ corresponding to the spin-dependent interaction. Adding these masses to the spin-averaged masses discussed in Section 2.1, one can find the mass $M=\bar{M}_{L}+\Delta M$ of a given DH tetraquark. In Tables (\ref{tab:cc1S}-\ref{tab:1P2}), we list the mass results for the double heavy tetraquarks in 1S and 1P waves.

\renewcommand{\tabcolsep}{0.35cm}
\renewcommand{\arraystretch}{1.2}
\begin{table}
\begin{center}
    \caption{Ground-state masses $M$ of the $cc$ tetraquarks, lowest threshold $T$ for its decaying into two D($c\bar{q}$) mesons and $\Delta=M-T$. All values are given in MeV.}
    \label{tab:cc1S}
    \begin{tabular}{ccccccccccc}
        \hline\hline
        System   & $IJ^{P}$  &$M$ & $T$   &$\Delta$  &\cite{Ebert:2007rn} &\cite{Braaten:2020nwp} &\cite{Eichten:2017ffp}&\cite{Zhang:2021yul}&\cite{Luo:2017eub}
        &\cite{LuCD:D20}\\ \hline
       {\multirow{4}{*}{$(cc)(\bar{u}\bar{d})$}}
        &$01^{+}$ &3997 &3871   &126 &$3935$ &$3947$ &$3978$ &       &&    \\
        &$10^{+}$ &4163 &3729   &434 &$4056$ &$4111$ &$4146$ &       &&    \\
        &$11^{+}$ &4185 &3871   &314 &$4079$ &$4133$ &$4167$ &$4117$ &$4201$&$4268$    \\
        &$12^{+}$ &4229 &4041   &188 &$4118$ &$4177$ &$4210$ &$4179$ &$4271$&$4318$      \\  \hline
       {\multirow{4}{*}{$(cc)(\bar{u}\bar{s})$}}
        &$\frac{1}{2}1^{+}$ &4180 &3975   &205 &$4143$ &$4124$ &$4156$ & &&    \\
        &$\frac{1}{2}0^{+}$ &4297 &3833   &464 &$4221$ &$4232$ &       & &&    \\
        &$\frac{1}{2}1^{+}$ &4315 &3975   &340 &$4239$ &$4254$ &       &$4314$ &$4363$ &4394   \\
        &$\frac{1}{2}2^{+}$ &4352 &4119   &233 &$4271$ &$4298$ &       &$4305$ &$4434$ &4440     \\  \hline
        {\multirow{3}{*}{$(cc)(\bar{s}\bar{s})$}}
        &$00^{+}$ &4420 &3936   &484 &$4359$ &$4352$ &       & &&    \\
        &$01^{+}$ &4436 &4080   &356 &$4375$ &$4374$ &       &$4382$ &$4526$&4493    \\
        &$02^{+}$ &4470 &4224   &246 &$4402$ &$4418$ &       &$4433$ &$4597$&4536    \\  \hline\hline
    \end{tabular}
    \end{center}
\end{table}

\renewcommand{\tabcolsep}{0.3cm}
\renewcommand{\arraystretch}{1.2}
\begin{table}
\begin{center}
    \caption{Ground-state masses $M$ of the $bb$ tetraquarks, lowest threshold $T$ for its decaying into two B($b\bar{q}$) mesons and $\Delta=M-T$. All values are in MeV. }
    \label{tab:bb1S}
    \begin{tabular}{ccccccccccc}
        \hline\hline
        System   & $IJ^{P}$  &$M$ & $T$   &$\Delta$  &\cite{Ebert:2007rn} &\cite{Braaten:2020nwp} &\cite{Eichten:2017ffp}&\cite{Zhang:2021yul}&\cite{Luo:2017eub}
        &\cite{LuCD:D20}\\ \hline
       {\multirow{4}{*}{$(bb)(\bar{u}\bar{d})$}}
        &$01^{+}$ &10530 & 10604  &-74 &$10502$ &$10471$ &$10482$ &       &&    \\
        &$10^{+}$ &10726 &10588   &168 &$10648$ &$10664$ &$10674$ &       &&    \\
        &$11^{+}$ &10733 &10604   &129 &$10657$ &$10671$ &$10681$ &$10854$ &$10875$&$10779$    \\
        &$12^{+}$ &10747 &10650   &97 &$10673$ &$10685$ &$10695$ &$10878$ &$10897$&$10799$      \\  \hline
       {\multirow{4}{*}{$(bb)(\bar{u}\bar{s})$}}
        &$\frac{1}{2}1^{+}$ &10713 &10693   &20 &$10706$ &$10644$ &$10643$ & &&    \\
        &$\frac{1}{2}0^{+}$ &10855 &10649   &208 &$10802$ &$10781$ &       & &&    \\
        &$\frac{1}{2}1^{+}$ &10861 &10693   &168 &$10809$ &$10788$ &       &$10974$ &$11010$ &10897   \\
        &$\frac{1}{2}2^{+}$ &10873 &10742  &131 &$10823$ &$10802$ &       &$10997$ &$11060$ &10915     \\  \hline
        {\multirow{3}{*}{$(bb)(\bar{s}\bar{s})$}}
        &$00^{+}$ &10976 &10739   &237 &$10932$ &$10898$ &       & &&    \\
        &$01^{+}$ &10981 &10786   &195 &$10939$ &$10905$ &       &$11099$ &$11199$&10986    \\
        &$02^{+}$ &10991 &10833   &158 &$10950$ &$10919$ &       &$11119$ &$11224$&11004    \\  \hline\hline
    \end{tabular}
    \end{center}
\end{table}

\begin{table}
\begin{center}
    \caption{Ground-state masses $M$ of $bc$ tetraquarks, lowest threshold $T$ for its decaying into B and D mesons and $\Delta=M-T$. All values are in MeV. }
    \label{tab:bc1S}
    \begin{tabular}{ccccccccccc}
        \hline\hline
        System   & $IJ^{P}$  &$M$ & $T$   &$\Delta$  &\cite{Ebert:2007rn} &\cite{Braaten:2020nwp} &\cite{Eichten:2017ffp}&\cite{Zhang:2021yul}&\cite{Luo:2017eub}
        &\cite{LuCD:D20}\\ \hline
       {\multirow{6}{*}{$(bc)(\bar{u}\bar{d})$}}
        &$00^{+}$ &7268 &7144   &154 &$7239$ &$7248$ &$7229$ &       &&    \\
        &$01^{+}$ &7269 &7190   &79 &$7246$ &$7289$ &$7272$ &       &&    \\
        &$11^{+}$ &7478 &7190   &288 &$7403$ &$7455$ &$7439$ &       &&    \\
        &$10^{+}$ &7458 &7114   &344 &$7383$ &$7460$ &$7461$ &       &&    \\
        &$11^{+}$ &7469 &7190   &279 &$7396$ &$7474$ &$7472$ & &&    \\
        &$12^{+}$ &7490 &7332   &158 &$7422$ &$7503$ &$7493$ &$7586$ &$7531$&$7582$      \\  \hline
       {\multirow{6}{*}{$(bc)(\bar{u}\bar{s})$}}
       &$\frac{1}{2}0^{+}$ &7451 &7232   &219 &$7444$ &$7422$ &$7406$ &       &&    \\
        &$\frac{1}{2}1^{+}$ &7452 &7277   &175 &$7451$ &$7456$ &$7445$ & &&    \\
        &$\frac{1}{2}1^{+}$ &7605 &7277   &328 &$7555$ &$7573$ & & &&    \\
        &$\frac{1}{2}0^{+}$ &7558 &7232   &356 &$7540$ &$7578$ &       & &&    \\
        &$\frac{1}{2}1^{+}$ &7579 &7277   &320 &$7552$ &$7592$ &       & & &  \\
        &$\frac{1}{2}2^{+}$ &7616 &7420   &196 &$7572$ &$7621$ &       &$7705$ & &     \\  \hline
        {\multirow{4}{*}{$(bc)(\bar{s}\bar{s})$}}
        &$01^{+}$ &7710 &7336   &374 &$7673$ &$4696$ &       & &&    \\
        &$01^{+}$ &7724 &7381   &343 &$7684$ &$7691$ &       & &&    \\
        &$10^{+}$ &7717 &7381   &336 &$7683$ &$7710$ &       & &&    \\
        &$11^{+}$ &7733 &7525   &208 &$7701$ &$7739$ &       &$7779$ &$7908$&7798    \\  \hline\hline
    \end{tabular}
    \end{center}
\end{table}

\renewcommand{\tabcolsep}{0.25cm}
\renewcommand{\arraystretch}{1.2}
\begin{table}[!htbp]
\begin{center}
    \caption{P wave masses (MeV) of the DH tetraquarks $T_{QQ}$ with configuration $QQ\{\bar{q}\bar{q}\}$.}
    \label{tab:1P1}
    \begin{tabular}{cccccccccc}
        \hline\hline
                        $J^{P}$ & $cc\{\bar{u}\bar{d}\}$  & $bb\{\bar{u}\bar{d}\}$ & $bc\{\bar{u}\bar{d}\}$ & $cc\{\bar{u}\bar{s}\}$  & $bb\{\bar{u}\bar{s}\}$
                         & $bc\{\bar{u}\bar{s}\}$  & $cc\{\bar{s}\bar{s}\}$  & $bb\{\bar{s}\bar{s}\}$ & $bc\{\bar{s}\bar{s}\}$\\ \hline
         $0^{-}$                  & 4429  & 10941 & 7687 & 4563  & 11110 & 7855 & 4696  & 11201 & 7959\\
         $1^{-}$                  & 4446  & 10946 & 7695 & 4579  & 11116 & 7863 & 4711  & 11206 & 7966\\
         $1^{-}$                  & 4447  & 10946 & 7696 & 4579  & 11116 & 7863 & 4710  & 11206 & 7966\\
         $1^{-}$                  & 4487  & 10959 & 7715 & 4613  & 11127 & 7879 & 4739  & 11215 & 7980\\
         $2^{-}$                  & 4484  & 10958 & 7714 & 4612  & 11126 & 7879 & 4741  & 11216 & 7981\\
         $2^{-}$                  & 4499  & 10963 & 7721 & 4624  & 11130 & 7884 & 4751  & 11219 & 7985\\
         $3^{-}$                  & 4517  & 10969 & 7730 & 4641  & 11136 & 7893 & 4766  & 11224 & 7993\\  \hline\hline
    \end{tabular}
    \end{center}
\end{table}
\renewcommand{\tabcolsep}{0.25cm}
\renewcommand{\arraystretch}{1.2}
\begin{table}[!htbp]
\begin{center}
    \caption{P wave masses (MeV) of the DH tetraquark with configurations $QQ[\bar{u}\bar{d}]$ and $QQ[\bar{u}\bar{s}]$. }
    \label{tab:1P2}
    \begin{tabular}{ccccccc}
        \hline\hline
        {\multirow{2}{*}{$J^{P}$}}   &\multicolumn{6}{c}{\rm Mass (MeV)} \\
        \cline{2-7}
                         & $cc[\bar{u}\bar{d}]$  & $bb[\bar{u}\bar{d}]$ & $bc[\bar{u}\bar{d}]$ & $cc[\bar{u}\bar{s}]$ & $bb[\bar{u}\bar{s}]$ & $bc[\bar{u}\bar{s}]$\\ \hline
         $0^{-}$         & 4253  & 10764 & 7520  & 4458  & 10983 & 7735\\
         $1^{-}$         & 4268  & 10769 & 7528  & 4472  & 10988 & 7741\\
         $2^{-}$         & 4298  & 10779 & 7542  & 4499  & 10996 & 7754\\ \hline\hline
    \end{tabular}
    \end{center}
\end{table}

It is seen from Tables (\ref{tab:cc1S}-\ref{tab:bc1S}) that our mass predictions of the DH tetraquarks are compatible with most other calculations cited. In the nonstrange sector, our prediction for the P wave masses are higher than the ground-state mass about 300 MeV for the $cc$ tetraquarks, about 300-400 MeV for the $bb$ tetraquarks and about 220 MeV for the $bc$ tetraquarks. In the strange sector, our prediction for the P wave masses are higher than the ground-state mass about 300 MeV for the $cc$ tetraquarks, about 270 MeV for the $bb$ tetraquarks and about 300 MeV for the $bc$ tetraquarks. It turns out that the monotonous suppression of the S-P wave splitting of tetraquark masses by heavy diquark mass $M_{QQ}$ breaks when light diquark becomes more strange.

Regarding our calculations on the spectra of DH tetraquarks, we make the
following remarks:

(i) The ground state masses of the DH tetraquarks increase mostly with the
quantum number $J^{P}$ of the states listed, as shown in Table \ref{tab:cc1S} through Table \ref{tab:bc1S}. A few
exceptions are the $IJ^{P}=01^{+}$ and $10^{+}$ states of the nonstrange
tetraquarks $(cc)(\bar{u}\bar{d})$ and $(bb)(\bar{u}\bar{d})$, the $%
IJ^{P}=(1/2)0^{+}$ and $(1/2)1^{+}$ states of the strange $bc$-tetraquarks $%
(bc)(\bar{u}\bar{s})$ and the $IJ^{P}=01^{+}$ and $10^{+}$ states of the
double-strange $bc$-tetraquarks $(bc)(\bar{s}\bar{s})$.

(ii) The state number of the excited DH tetraquarks are far less than that
allowed by the four-body quark systems. As such, some P-wave excited states
have not been predicted in Tables \ref{tab:1P1} and \ref{tab:1P2}, which are associated with the internal excitations of the light or heavy diquarks.

(iii) Generally, a multiquark state may form a mixture of all possible color
components, including the color singlets and hidden color configurations.
For example, the $T_{cc}^{+}$ state $|\bar{3}_{c},3_{c}\rangle $ (with color
structure $\bar{3}_{c}\times 3_{c}$) with $IJ^{P}=01^{+}$ can in principle
mix with the tetraquark $|6_{c},\bar{6}_{c}\rangle$ (color structure $6_{c}\times \bar{6}_{c}$) with $IJ^{P}=01^{+}$ to form the two states $|1_{c},1_{c}\rangle$ and $|8_{c},8_{c}\rangle$:
\begin{equation}
|1_{c},1_{c}\rangle =-\frac{1}{\sqrt{3}}|\bar{3}_{c},3_{c}\rangle + \sqrt{\frac{2}{3}} |6_{c},\bar{6}_{c}\rangle,    \label{11c}%
\end{equation}
\begin{equation}              \label{88c}
|8_{c},8_{c}\rangle =\sqrt{\frac{2}{3}}|\bar{3}_{c},3_{c}\rangle + \frac{1}{\sqrt{3}} |6_{c},\bar{6}_{c}\rangle.
\end{equation}

Other possibility is that the $T_{cc}^{+}$ state is mixture of the two color
singlets (molecule states) $DD^{\ast }$ and $D^{\ast }D^{\ast }$, and two
color octets $D_{8}D^{*}_{8}$ and $D^{*}_{8}D^{*}_{8}$. These possible mixture effects as well as the coupling with the molecule channels of the $DD^{\ast }$ and $D^{\ast }D^{\ast }$ may
shift the ground-state mass prediction $3997$ MeV of the state $|\bar{3}%
_{c},3_{c}\rangle $ (Table \ref{tab:cc1S}) down to be near to the measured mass (about $3874.7 $ MeV).

\section{Summary and Remarks}
In this work, we utilize linear Regge relation, which is derived from QCD string model, and mass scaling with respect to the $D(D_{s})$ mesons to study low-lying excited mass spectra of the doubly heavy tetraquark in heavy diquark-antidiquark picture. For the spin-independent masses of the DH tetraquarks, the measured data of the $\Xi_{cc}$ baryon and other compatible estimates of ground-state masses of doubly heavy baryons are used to extract the trajectory parameters, including the heavy diquark $M_{QQ}$ ($Q=c,b$). For the spin-dependent masses, we employ the $D/D_{s}$ meson masses in the relation of mass scaling to determine the hyperfine mass splitting. We find that except for the ground states of doubly bottom tetraquarks $T_{bb}$($bb\bar{u}\bar{d}$ $1S$ with $ IJ^P=01^+$), all DH tetraquarks are not stable against strong decays, provided that the DH tetraquarks are pure state of compact exotic hadrons.

The heavy quark pair in diquark $QQ$ in $\bar{3}_{c}$ may (when $M_{Q}$ is very large) stay close to each other to form a compact core due to the
strong Coulomb interaction, with the light quarks moving around the $QQ$%
-core\cite{Luo:2017eub}, while it is also possible (when $M_{Q}$ is comparable with $1/\Lambda _{QCD} \approx 3$ GeV$^{-1}$ ) that $Q$ attracts $\bar{q}$ to bind
into a colorless clustering (in $1_{c}$) and other pair of $Q^{\prime }$ and $%
\bar{q}^{\prime }$ binds into another colorless clustering (in $1_{c}$),
giving a molecular system $(Q\bar{q})(Q^{\prime }\bar{q}^{\prime })$. In the
former case, DH tetraquarks mimic the helium-like QCD-atom, while in the
later case, DH tetraquarks resemble the hydrogen-like QCD molecules. In the
real world with finite heavy-quark mass, especially in the case of the $cc$
tetraquark state, a DH tetraquark may form exotic states of compact
tetraquark, or form an exotic molecular state consisting of two heavy mesons,
or mixture of the both structures\cite{DZ22}.

We stress that our picture of anitdiquark-heavy-diquark relies on the
diquark size $\langle R\rangle $ compared to the scale $1/\Lambda _{QCD}$, as two heavy quarks $Q$ and $Q^{\prime }$ in color
antitriplet ($\bar{3}_{c}$) in DH tetraquark may not act as a compact color
source when $\langle R\rangle \gg 1/\Lambda _{QCD}$. Explicit size of
diquark depends on two-quark dynamics\cite{BW13,GLP19,KN13} and the mixing of the
molecule components in DH tetraquarks is of interest and remains to be
explored.

\textbf{ACKNOWLEDGMENTS}

D.J. is supported by the National Natural Science Foundation of China under the No. 12165017. Y.S thanks Wen-Xuan Zhang for many discussions.

\section*{\textbf{Appendix} \textbf{A}}
\setcounter{equation}{0}
\renewcommand{\theequation}{A\arabic{equation}}
For a system composed of a heavy-diquark $QQ$ and light antidiquark $\bar{q}\bar{q}$, the matrix elements of $\mathbf{L} \cdot \mathbf{S}_{qq}$, $\mathbf{L} \cdot \mathbf{S}_{Q Q}$,$\mathbf{\widehat{B}}$
and $\mathbf{S}_{Q Q} \cdot \mathbf{S}_{qq}$ may be evaluated by explicit construction of the tetraquark states with a given $J_{3}$ as linear combinations of states $\left|S_{Q Q3}, S_{qq 3}, L_{3}\right\rangle$, where
$S_{qq 3}+S_{Q Q 3}+L_{3}=S_{\bar D3}+S_{D3}+L_{3}=J_{3}$.  Due to the rotation invariance of the matrix elements, it suffices to use a single $J_{3}$ for each and, one can use
\begin{equation}\label{LS}
    \mathbf{L} \cdot \mathbf{S}_{i}=\frac{1}{2}\left[L_{+} S_{i-}+L_{-} S_{i+}\right]+L_{3} S_{i 3},
\end{equation}
to find their elements by applying $\mathbf{L} \cdot \mathbf{S}_{i}$ on the third components of angular momenta, where $i=D, \bar D$.
In the $\left|J, J_{3}\right\rangle$ representation in the $LS$ coupling, these
component are given by the following basis states(when $S_{D}=S_{\bar D}=1$ )
\begin{equation}\label{a}
    \begin{aligned}
\left|{ }^{3} P_{0}, J=0\right\rangle &=\frac{1}{\sqrt{6}}|1,-1,0\rangle+\frac{1}{\sqrt{6}}|0,1,-1\rangle-\frac{1}{\sqrt{6}}|-1,1,0\rangle \\
&-\frac{1}{\sqrt{6}}|1,0,-1\rangle-\frac{1}{\sqrt{6}}|0,-1,1\rangle+\frac{1}{\sqrt{6}}|-1,0,1\rangle,
\end{aligned}
\end{equation}
\begin{equation}\label{b}
   \left|{ }^{1} P_{1}, J=1\right\rangle=\frac{1}{\sqrt{3}}|1,-1,1\rangle-\frac{1}{\sqrt{3}}|0,0,1\rangle+\frac{1}{\sqrt{3}}|-1,1,0\rangle,
\end{equation}
\begin{equation}\label{c}
\left|{ }^{3} P_{1}, J=1\right\rangle=\frac{1}{2}|1,-1,1\rangle-\frac{1}{2}|-1,1,1\rangle-\frac{1}{2}|1,0,0\rangle+\frac{1}{2}|0,1,0\rangle,
\end{equation}
\begin{equation}\label{d}
\begin{aligned}
\left|{ }^{5} P_{1}, J=1\right\rangle &=\frac{1}{2 \sqrt{15}}|1,-1,1\rangle+\frac{1}{\sqrt{15}}|0,0,1\rangle+\frac{1}{2 \sqrt{15}}|-1,1,1\rangle \\
&-\frac{1}{2} \sqrt{\frac{3}{5}}|1,0,0\rangle-\frac{1}{2} \sqrt{\frac{3}{5}}|0,1,0\rangle+\sqrt{\frac{3}{5}}|1,1,-1\rangle,
\end{aligned}
\end{equation}
\begin{equation}\label{e}
    \left|{ }^{3} P_{2}, J=2\right\rangle=\frac{1}{\sqrt{2}}|1,0,-1\rangle+\frac{1}{\sqrt{2}}|0,1,1\rangle,
\end{equation}
\begin{equation}\label{f}
    \left|{ }^{5} P_{2}, J=2\right\rangle=\frac{1}{\sqrt{6}}|1,0,1\rangle+\frac{1}{\sqrt{6}}|0,1,1\rangle-\sqrt{\frac{2}{3}}|1,1,0\rangle,
\end{equation}
\begin{equation}\label{g}
    \left|{ }^{5} P_{3}, J=3\right\rangle=|1,1,1\rangle,
\end{equation}

In the state subspace of $J=1$, using the basis $\left[{ }^{1} P_{1},{ }^{1} P_{1}\right]$, $\left[{ }^{1} P_{1},{ }^{3} P_{1}\right]$,
 $\left[{ }^{1} P_{1},{ }^{5} P_{1}\right]$ and $\left[{ }^{3} P_{1},{ }^{3} P_{1}\right]$, $\left[{ }^{3} P_{1},{ }^{5} P_{1}\right]$, $\left[{ }^{5} P_{1},{ }^{5}P_{1}\right]$, one can compute the matrix elements of $\mathbf{L} \cdot \mathbf{S}_{i}$, $\mathbf{B}$ and $\mathbf{S}_{\bar D} \cdot \mathbf{S}_{D}$,
\begin{equation}\label{h}
\left\langle\mathbf{L} \cdot \mathbf{S}_{\bar{D}}\right\rangle_{J=1}=\left[\begin{array}{ccc}
0 & \frac{2}{\sqrt{3}} & 0 \\
\frac{2}{\sqrt{3}} & -\frac{1}{2} & \frac{\sqrt{15}}{6} \\
0 & \frac{\sqrt{15}}{6} & -\frac{3}{2}
\end{array}\right],\left\langle\mathbf{L} \cdot \mathbf{S}_{D}\right\rangle_{J=1}=\left[\begin{array}{ccc}
0 & -\frac{2}{\sqrt{3}} & 0 \\
-\frac{2}{\sqrt{3}} & -\frac{1}{2} & -\frac{\sqrt{15}}{6} \\
0 & -\frac{\sqrt{15}}{6} & -\frac{3}{2}
\end{array}\right],
\end{equation}
\begin{equation}\label{i}
\left\langle\mathbf{S}_{\bar{D}} \cdot \mathbf{S}_{D}\right\rangle_{J=1}=\left[\begin{array}{ccc}
-2 & 0 & \frac{1}{6 \sqrt{5}} \\
0 & -1 & 0 \\
\frac{1}{6 \sqrt{5}} & 0 & 1
\end{array}\right],\langle \mathbf B\rangle_{J=1}=\left[\begin{array}{ccc}
0 & 0 & \frac{32}{15 \sqrt{5}} \\
0 & 1 & 0 \\
\frac{32}{15 \sqrt{5}} & 0 & -\frac{7}{5}
\end{array}\right],
\end{equation}
In the subspace of $J=2$, one finds
\begin{equation}\label{j}
\left\langle\mathbf{L} \cdot \mathbf{S}_{\bar{D}}\right\rangle_{J=2}=\left[\begin{array}{cc}
\frac{1}{2} & \frac{\sqrt{3}}{2} \\
\frac{\sqrt{3}}{2} & -\frac{1}{2}
\end{array}\right],\left\langle\mathbf{L} \cdot \mathbf{S}_{D}\right\rangle_{J=2}=\left[\begin{array}{cc}
\frac{1}{2} & -\frac{\sqrt{3}}{2} \\
-\frac{\sqrt{3}}{2} & -\frac{1}{2}
\end{array}\right],
\end{equation}
\begin{equation}\label{k}
\left\langle\mathbf{S}_{\bar{D}} \cdot \mathbf{S}_{D}\right\rangle_{J=2}=\left[\begin{array}{cc}-1 & 0 \\ 0 & 1\end{array}\right] ,
\langle \mathbf B\rangle_{J=2}=\left[\begin{array}{cc}
-\frac{1}{5} & 0 \\
0 & \frac{7}{5}
\end{array}\right],
\end{equation}
In the subspace of $J=0$, the matrix elements are
\begin{equation}\label{m}
\begin{array}{l}
\langle\mathbf{L} \cdot \mathbf{S}_{}\rangle_{J=0}=-1,\langle\mathbf{L} \cdot \mathbf{S}_{\bar{D}}\rangle_{J=0}=-1, \\
\left\langle\mathbf{S}_{\bar{D}} \cdot \mathbf{S}_{D}\right\rangle_{J=0}=-1, \quad\langle\mathbf{B}\rangle_{J=0}=-2,
\end{array}
\end{equation}
In the subspace of $J=3$, the results are
\begin{equation}\label{n}
\begin{array}{l}
\langle\mathbf{L} \cdot \mathbf{S}_{}\rangle_{J=3}=1,\langle\mathbf{L} \cdot \mathbf{S}_{\bar{D}}\rangle_{J=3}=1, \\
\left\langle\mathbf{S}_{\bar{D}} \cdot \mathbf{S}_{D}\right\rangle_{J=3}=1, \quad\langle\mathbf{B}\rangle_{J=3}=-\frac{2}{5},
\end{array}
\end{equation}

Given the above matrices, one can solve eigenvalues $\lambda$ of
$\mathbf{L}\cdot\mathbf{S}_{qq}$ and the corresponding eigenvectors for a given
$J$, and for that $J$ one can write the hadron states $\left\vert
J,j\right\rangle $ in the $jj$ coupling. Then, these hadron states can be expressed to be the linear combinations of $LS$ bases $\left\vert ^{2S+1}P_{J}\right\rangle $:
\begin{equation}\label{s}
\lambda=+1:|J=0, j=0\rangle=\left|1{ }^{3} P_{0}\right\rangle,
\end{equation}
\begin{equation}\label{y}
\lambda=-2:|J=1, j=0\rangle=\frac{1}{3}\left|1^{1} P_{1}\right\rangle-\frac{1}{\sqrt{3}}\left|1^{3} P_{1}\right\rangle+\frac{\sqrt{5}}{3}\left|1^{1} P_{1}\right\rangle,
\end{equation}
\begin{equation}\label{x}
\lambda=-1:|J=1, j=1\rangle=\frac{1}{\sqrt{3}}\left|1^{1} P_{1}\right\rangle-\frac{1}{2}\left|1^{3} P_{1}\right\rangle-\frac{1}{2} \sqrt{\frac{5}{3}}\left|1^{1} P_{1}\right\rangle,
\end{equation}
\begin{equation}\label{o}
\lambda=1:|J=1, j=2\rangle=\frac{\sqrt{5}}{3}\left|1^{1} P_{1}\right\rangle-\frac{1}{2} \sqrt{\frac{5}{3}}\left|1^{3} P_{1}\right\rangle+\frac{1}{6}\left|1^{1} P_{1}\right\rangle,
\end{equation}
\begin{equation}\label{v}
\lambda=-1:|J=2, j=1\rangle=\frac{1}{2}\left|1^{3} P_{2}\right\rangle-\frac{\sqrt{3}}{2}\left|1^{5} P_{2}\right\rangle,
\end{equation}
\begin{equation}\label{e}
\lambda=1:|J=2, j=2\rangle=\frac{\sqrt{3}}{2}\left|1^{3} P_{2}\right\rangle-\frac{1}{2}\left|1^{5} P_{2}\right\rangle,
\end{equation}
\begin{equation}\label{j}
\lambda=+1:|J=3, j=2\rangle=\left|1^{5} P_{3}\right\rangle.
\end{equation}
This gives the required baryon states in the heavy-diquark limit, by which the
diagonal matrix elements of $\mathbf{L}\cdot\mathbf{S}_{QQ}$,
$\mathbf{\widehat{B}}$ and $\mathbf{S}_{QQ}\cdot\mathbf{S}_{qq}$ can be obtained.

In the $\left|J, J_{3}\right\rangle$ representation in the $LS$ coupling, these
component are given by the following basis states(for $S_{D}=1$,$S_{\bar D}=0$)
\begin{equation}\label{S}
    \begin{array}{l}
\left|{ }^{3} P_{0}, J=0\right\rangle=\frac{1}{\sqrt{3}}|0,1,-1\rangle+\frac{1}{\sqrt{3}}|0,-1,1\rangle-\frac{1}{\sqrt{3}}|0,0,0\rangle, \\
\left|{ }^{3} P_{0}, J=1\right\rangle=\frac{1}{\sqrt{2}}|0,1,0\rangle-\frac{1}{\sqrt{2}}|0,0,1\rangle, \\
\left|{ }^{3} P_{2}, J=2\right\rangle=|0,1,1\rangle,
\end{array}
\end{equation}
In the following, we write only the nonzero matrix elements for the spin-orbit coupling operators with $J=0, 1$:
\begin{equation}\label{Y}
   \begin{array}{l}
\langle\mathbf{L} \cdot \mathbf{S}_{\bar{D}}\rangle_{J=0}=-2 (  J=0) \\
\end{array}
\end{equation}
\begin{equation}\label{X}
   \begin{array}{l}
\langle\mathbf{L} \cdot \mathbf{S}_{\bar{D}}\rangle_{J=1}=-1 (  J=1) \\
\end{array}
\end{equation}
\begin{equation}\label{LJ}
   \begin{array}{l}
\langle\mathbf{L} \cdot \mathbf{S}_{\bar{D}}\rangle_{J=2}=1, (  J=1). \\
\end{array}
\end{equation}
with the results collected in Table \ref{tab:matrix elements}.

\end{document}